Software Development Kits and Wearable Devices in Physical Activity Research

Jason Tsang, Harry Prapavessis

Western University, School of Kinesiology

April 10th, 2023

*Author's Notes*

Tsang – Development of the application, Study Design, Creation of Research Question,

Authoring Final Manuscript

Prapavessis – Creation of Research Question, Study Design, Supervision

Table of Contents

Abstract

Introduction: The Canadian Guidelines recommend physical activity for overall health benefits, including cognitive, emotional, functional, and physical health. However, traditional research methods are inefficient and outdated. This paper aims to guide researchers in enhancing their research methods using software development kits and wearable smart devices.

Methods: A generic model application was transformed into a research-based mobile application based on the UCLA researchers who collaborated with Apple. First, the research question and goals were identified. Then, three open-source software development kits (SDKs) were used to modify the generic model into the desired application. ResearchKit was used for informed consent, surveys, and active tasks. CareKit was the protocol manager to create participant protocols and track progress. Finally, HealthKit was used to access and share health-related data. The content expert evaluated the application, and the participant experience was optimized for easy use. The collected health-related data were analyzed to identify any significant findings.

Results: Wearable health devices offer a convenient and non-invasive way to monitor and track health-related information.

Conclusion: Leveraging the data provided by wearable devices, researchers can gain insights into the effectiveness of interventions and inform the development of evidence-based physical activity guidelines. The use of software development kits and wearable devices can enhance research methods and provide valuable insights into overall health benefits.

Software Development Kits and Wearable Devices in Physical Activity Research

## Introduction

The Canadian Guidelines recommend physical activity as a means of maintaining overall health. Physical activity (PA) is a well-established behaviour that contributes to cognitive, emotional, functional, and physical health. In addition to physical activity, evidence suggests the importance of minimizing sedentary time in maintaining overall health. However, understanding the relationship between physical activity, sedentary behaviour, and health is more complex. Most of the evidence on physical activity and health comes from studies that have used self-reported measures, which are only moderately correlated with objective criteria and typically overestimate physical activity levels.

These studies have been performed using traditional, inefficient, and outdated methods. Therefore, it is important to continue researching and finding new approaches to understanding the relationship between physical activity, sedentary behaviour, and health. Previous studies on physical activity and health have been performed using traditional, inefficient, and outdated methods such as self-reported measures (Weatherson et al., 2021). These methods are subject to limitations such as time, geographic, and financial constraints, as well as potential participant bias. Additionally, these methods limit recruitment's scalability, transferability, and scope.

## Background

The number of studies that contained app usage were mostly "Business and Retail" related comprising of 40% of all examined primary studies. "Health and Medical" related research containing app usage comprised of 9.09% of all studies (Knight et al., 2015; Shamsujjoha et al., 2021). This demonstrates a gap in understanding between researchers and the rapid developments in wearable smart devices (WSD) capabilities. There have only been a handful of noteworthy adoptions of such technologies in the research sector although such technologies have become ubiquitous in nearly every aspect of daily life (Jayatilleke et al., 2018; Knight et al., 2015).

The feasibility of using two devices to measure sedentary time and physical activity has been demonstrated previously in a community-based cohort of older adults (Rosenberg et al., 2020), highlighting the potential for using wearable technology in physical activity research to overcome the limitations of traditional methods.

Therefore, this paper intends to guide researchers on how to apply and enhance their existing research methods with software development kits and wearable smart devices. With recommendations and best practices addressed with more human-centric aspects in app development.

## Methodology

### Wearable Smart Devices

Smart wearable health devices are electronic devices that are worn on the body, often as accessories, and are designed to monitor and track health-related information. Many wearable devices are equipped with sensors that can measure physical activity, including steps taken, distance traveled, and calories burned. They may also be able to track heart rate, sleep patterns, and other health metrics. These devices are often designed to sync with a mobile app or other software platform that can provide detailed analysis of the collected data (Apple ®, 2022, 2022; Bai et al., 2021).

Researchers can use smart wearable health devices to measure physical activity levels in a variety of settings and populations, including community-based interventions (Bai et al., 2021). Wearable devices can track the duration and intensity of physical activity, as well as provide information on sedentary behavior, which has been linked to adverse health outcomes (Abt et al., 2018; Bai et al., 2021; Falter et al., 2019; Khushhal et al., 2017).

As of 2023 Apple Watches, Garmin devices, and Fitbits are popular wearable devices that can be used for physical activity research (Rosenberg et al., 2020). There have been many validation studies conducted on these devices to assess their accuracy and validity for measuring physical activity (Abt et al., 2018; Khushhal et al., 2017; Werner et al., 2023).

**Software Development Kits**

A Software Development Kit (SDK) is a collection of software development tools that are bundled together to help developers create software applications for a specific platform, operating system, or framework. The SDK typically includes libraries, documentation, and examples to help developers get started with building applications for the targeted platform. SDKs can also help ensure that the resulting applications are compatible with other software and systems that are commonly used in research. This can help ensure that the applications are reliable, secure, and can be easily integrated with other tools and systems. SDKs are an important tool for medical researchers who are developing software applications for data analysis and processing. By providing the necessary tools and resources, SDKs can help researchers create powerful and effective applications without having to start from scratch.

Using appropriate SDKs can be used to demystify the app development process, reduce complexity, provide development principles, helping to achieve maximal scalable methods, productivity, and cost-reduction.

High quality and well-maintained SDKs often include application programming interfaces (APIs), documentation for such APIs, libraries of pre-built lines of code, and testing tools.

### Application Programming Interfaces

Application Programming Interfaces (APIs) or colloquial known as libraries are a collection of pre-written code that developers can use to create applications that interface with the platform or programming language. These libraries may include functions, modules, and classes that provide a specific set of features and functionality.

### Documentation

Documentation should provide comprehensive guidance to developers on how to use the SDK's components to build applications. This documentation may include code samples, tutorials, and other reference material.

### Integrated Development Environment (IDE)

An IDE is a software application that provides a comprehensive environment for developers to write, test, and debug their code. It typically includes a code editor, debugger, and other development tools.

## Model Driven Development

Model-driven development (MDD) is a software development approach that emphasizes creating models to generate code automatically. In app development, MDD involves using these models to describe the app's functionality and behavior, then automatically generating code using specialized tools. An overview of the model-driven development process is presented in Table 1.

MDD can be used to streamline health research app development and ensure that the app meets requirements. By using models to capture app specifications, such as use cases, domain models, and process models, MDD can provide a structured approach that enhances research efficiency and accuracy.

After creating the models, MDD tools generate the app's code, including user interface, data management, and other essential functions. The generated code is then tested and refined to meet app requirements and specifications.

MDD offers numerous benefits for health research app development, including productivity improvement, reduced development time, and increased app quality and accuracy. Furthermore, it can ensure that the app complies with ethical and regulatory standards, critical aspects of health research.

These models include (1) Use case Models: To describe the various use cases or scenarios that the app will support. (2) Domain Models: Capture the key concepts and entities in the health research domain, such as patients, medical conditions, and treatments. (3) Process Models: Describe the workflows and processes that the app will support, such as data collection, analysis, and reporting (Table 2).

**Stakeholders Framework**

In this study, five groups of stakeholders who are involved in the development and evaluation of the mobile application are identified. These groups include content experts who validate the content, novice users who evaluate the overall effectiveness of the application, developers who build the application and test its ease of use, and researchers who analyze the impact of the innovation (see Table 3). By involving these different stakeholders, we aim to ensure that the application is developed and evaluated comprehensively from different perspectives, leading to a more effective and user-friendly educational tool.

**Development Phases**

This guide provides an example of how to develop a similar application using an existing study and breaks down the steps into three phases based on the MDD approach. The UCLA

"Digital Mental Health Study" in collaboration with Apple was used as the basis for transforming a generic model application into a research-based mobile application.

**Phase I**

To commence the development of a research-based application, it is crucial to identify the study's objective and the primary functionalities of the application. For instance, in the case of the "Digital Mental Health Study," the primary objective was to leverage objective health-related measures collected on smart devices to advance knowledge on the causes and trajectories of depression, stress, anxiety, and their potential relationship to other medical conditions such as cardiovascular disease, diabetes, neurological disorders, and cancer. This involved exploring and developing objective measures relevant to these conditions. ResearchKit and CareKit can be utilized to create the necessary features to support these objectives, such as data collection tools and tracking functionalities.

**Phase II**

Next, three open-source software development kits (SDKs) were utilized to tailor the generic model into the desired application. ResearchKit was leveraged to obtain informed consent, conduct surveys, and execute active tasks. CareKit served as the protocol manager, enabling the creation of participant protocols and the tracking of their progress. Lastly, HealthKit was employed to gain access to and share health-related data.

*Eligibility*

To ensure efficient screening of participants, it is important to determine eligibility promptly. The eligibility section should be placed after the introduction and before the consent section, so participants who are not eligible for the study need not view the consent section. It is recommended to present only the necessary eligibility requirements in simple and straightforward language, making it easy for participants to enter their information as seen in figure 1. This approach can streamline the participant selection process and enhance the overall efficiency of the study.

### Informed Consent

The participants were provided with the letter of information/consent (LOI/C) directly on the iOS application. They were also given an option to share a pdf copy. The iOS developer license agreement and health-related research requirements were taken into consideration. Since the requirements from the license agreement were in line with standard research ethics guidelines, no additional changes were made to the LOI/C. The LOI/C contained information about the research's nature, purpose, and duration, procedures, risks, and benefits to the participant, confidentiality, and data handling (including sharing with third parties), a contact point for participant questions, and the withdrawal process.

The onboarding step utilized Research Kit's survey module to provide the LOI/C and obtain consent through a digital signature, name, and the date of survey completion. Each component of the LOI/C was visually displayed directly on the application during the study's onboarding step, accompanied by both visual and written means to minimize potential misunderstandings. At the end of the onboarding step, a PDF version of the comprehensive LOI/C was displayed, followed by a form to collect the name and digital signature. The figure 2 provides a visual reference of the sequential display of the LOI/C and its accompanying form.

After obtaining consent, participants were requested to disclose and provide permission for all the health-related data collected, as displayed in figure 3.

### Surveys

The Beck Depression Index II (BDI-II), Godin Leisure Physical Activity Questionnaire (GLPQ) and the Pittsburgh Sleep Quality index (PSQI) were used as stand-in examples in the surveys. To create a depression questionnaire, self-reported physical activity, and sleep questionnaire using ResearchKit, one needs to follow certain steps. The task steps should include an instruction step, a question step, and a summary step.

In the question step, BDI-II questions can be added as text questions with an answer format defined as a scale from 0 to 3. Skip patterns can also be added based on the participant's answers. Similarly, for the Godin Leisure questions, they can be added as text questions with an answer format defined as a numeric value and skip patterns can also be added based on the participant's answers. Lastly, the PSQI questions can be added as text questions with an answer format defined as a scale from 0 to 3. Skip patterns can also be added based on the participant's answers. Sample visualisation of the questionnaires are displayed in figure 4.

By using ResearchKit to create these questionnaires, researchers can collect a vast amount of data remotely and in real-time, providing them with valuable insights into participants' health and well-being. Additionally, participants can conveniently complete the questionnaires using their iOS devices, increasing accessibility and convenience.

### Point of Contact

In utilizing CareKit for a health research study, it is important to include contact information for various parties involved in the study, including the researchers, ethics board, and academic institution. To create contact information in CareKit, one can first define the required fields such as name, phone number, email, and address. These fields should be presented in a clear and organized manner for easy reference by participants. Additionally, one can incorporate features such as a "tap-to-call" button for phone numbers and a "tap-to-email" button for email addresses to enhance the user experience. CareKit also allows for customization of the contact screen, such as adding logos or images related to the study. By providing easily accessible and organized contact information, participants can quickly and efficiently reach out to the necessary parties for any questions or concerns related to the study as shown in figure 5.

**Phase III**

*User Experience*

ResearchKit and CareKit provide tools that can enhance the user experience in research and health apps. These tools enable researchers to create mobile apps that are user-friendly, intuitive, and accessible. By utilizing these tools, researchers can optimize the participant experience and improve engagement and retention in studies.

For example, ResearchKit provides customizable templates for creating study modules that can be easily integrated into a mobile app. These modules have a consistent user interface, making it easy for participants to navigate through the app and complete tasks. Additionally, ResearchKit includes built-in features such as customizable forms and interactive questionnaires, which can help to streamline the data collection process and improve participant comprehension.

Similarly, CareKit provides a framework for health management and allows for the creation of customized user interfaces that can be tailored to meet the unique needs of participant. This can include features such as reminders, and progress tracking, which can help participants to stay engaged and motivated in their health management demonstrated in figure 6.

Overall, the use of ResearchKit and CareKit can greatly enhance the user experience in research and health apps, leading to improved engagement, adherence, and overall participant satisfaction.

### Data Management

The ResearchKit framework provided a profile screen that allowed users to manage their personal information while they were in a research app. This feature provided a centralized location for participants to store and update their personal information, including demographics, medical history, and contact information. In addition, a screen was created to motivate users and provide them with a way to track their progress in the study. Participants were able to access both areas at any time throughout the study (Figure 6).

### Data Visualization

Displaying summary statistics, graphs, or charts can provide a valuable way to visually represent participants' progress in a research study. ResearchKit and CareKit offer tools for creating these visualizations and presenting them to participants. However, it is important to carefully consider whether it is appropriate to show participants their data in real-time based on the goals and design of the study. In some cases, real-time feedback may be desirable and help motivate participants to continue with the study. In other cases, it may be more appropriate to wait until the study is complete before presenting participants with their data. Regardless of the timing, the use of visualizations can provide a powerful way to engage participants and help them better understand their progress in the study. In this example data visualization was omitted as all the data and its visualization was provided in the health application.

## Results

## Researcher

Researchers can benefit from these SDKs in several ways. Firstly, these tools can streamline the process of developing mobile health applications for research purposes. This can save researchers time and resources, allowing them to focus on the scientific aspects of their research. Secondly, these tools can help researchers collect and manage large amounts of health-related data, including data related to physical activity. This data can be used to gain insights into health and disease, as well as to inform the development of new interventions and treatments (Figure 6). Thirdly, these tools can improve the quality and accuracy of health-related data by minimizing errors and reducing the risk of data loss. Finally, these tools can help researchers comply with regulatory requirements and ethical standards related to health research. Overall, these SDKs can help researchers conduct high-quality research that is both efficient and effective.

Wearable smart devices such as Apple Watches can be beneficial for researchers in physical activity research as they provide a convenient and non-invasive method for collecting continuous and objective data related to physical activity and other health-related metrics. This data can be used to gain insights into participants' behavior and activity patterns, as well as to monitor the effectiveness of interventions aimed at promoting physical activity. Additionally, the use of wearable devices can increase the scale of participant recruitment and reduce the cost of data collection compared to traditional methods, making it a valuable tool for researchers in physical activity research.

**Content Expert**

Content experts can benefit from the use of these SDKs in several ways. Firstly, by using these tools, content experts can focus on the content and user experience of the app, without having to worry about the technical details of the app development process. Secondly, content experts can transform relevant information into a digital and mobile-friendly format, while considering the user experience of the participant. This enables information to be displayed in a manner that is easy to understand. Thirdly, these SDKs provide pre-built modules and components that can be used to speed up the app development process, thereby reducing the overall development time and cost. Finally, the data collected through these apps can be used by content experts to gain insights into user behavior, preferences, and needs, which can be used to improve the app and overall user experience.

**Software Developer**

Software developers can benefit from using these SDKs in several ways. First, these tools provide pre-built and standardized components for various functionalities of a health research app, such as data collection, consent management, and data storage. This can save software engineers significant development time and effort, allowing them to focus on other aspects of the app.

Second, these SDKs are designed to be easy to use, even for software engineers with limited experience in mobile app development. They come with comprehensive documentation and sample code, making it easier to get started and integrate the SDKs into an app.

Third, these SDKs provide access to a large community of developers who use and contribute to the SDKs. This can be helpful in troubleshooting issues and finding solutions to problems, as well as sharing best practices and tips for app development.

Finally, using SDKs for data collection, storage, and management can help ensure that the app meets regulatory and ethical standards, which is critical in the context of health research.

**Participant**

Participants can benefit from the use of these SDKs in several ways. Firstly, the use of mobile devices and wearables for data collection offers a convenient and non-invasive way to monitor and track health-related information, which can lead to better awareness and management of their own health. Additionally, the use of these tools can enable participants to contribute to research studies and help advance scientific knowledge, without the need for them to physically attend a research facility. The use of informed consent processes and data sharing options can also help ensure that participants are fully informed and have control over their participation in research. Finally, the data collected from participants can be used to identify individual health trends, which can help guide personalized interventions or treatments.

The findings of this study are subject to several limitations and challenges. Firstly, the limited support structures for inventors can limit their ability to develop and test new ideas, which can hinder progress in the field. Secondly, the lack of structures for sustaining technological interventions can result in innovations being short-lived and not having a lasting impact. Thirdly, the lack of time for academics to modify and verify content for mobile applications can slow down the development process. Additionally, the absence of adequate technological infrastructure can make it difficult to develop and implement mobile systems effectively. Moreover, the high development costs associated with developing and implementing mobile systems can limit their widespread adoption and use. Furthermore, the traditional institutional ethics regulation can slow down the approval process for new study protocols, which can limit the ability to conduct research in certain areas. The considerable time needed to carry out usability testing and modify errors can delay the implementation of new technologies and systems. The need for induction training for students to use mobile technologies can create a barrier to their adoption and use. Lastly, the need for technical support throughout the learning process can be challenging to provide and can affect the effectiveness of the learning experience. These limitations/challenges can be used to highlight the difficulties and barriers associated with the development and implementation of new technological innovations and can provide direction for future research and development in the field.

**Conclusion**

In conclusion, the development of mobile health research applications using open-source SDKs such as ResearchKit, HealthKit, and CareKit offers numerous benefits for researchers, content experts, participants, and software engineers. MDD provides a structured approach to app development that can improve the efficiency and accuracy of the research process, while the SDKs offer powerful tools for collecting, managing, and analyzing health-related data. Researchers can benefit from increased scale of participant recruitment and remote data collection. Content experts can transform relevant information into a digital and mobile-friendly format. Participants can benefit from increased convenience and personalized health monitoring. Software developer can benefit from reusable components and improved code quality. The use of MDD and open-source SDKs has the potential to revolutionize the way health research is conducted, offering new opportunities for data-driven insights into the effectiveness of interventions and inform the development of evidence-based physical activity guidelines. The use of software development kits and wearable devices can enhance research methods and provide valuable insights into overall health benefits. Overall, the integration of these technologies allows for more comprehensive and convenient data collection, ultimately leading to a better understanding of physical activity and overall health.

Tables

Table 1

*Model Driven Development Stages.*

| Stage | Description |
| --- | --- |
| Requirements Gathering | Gathering and documenting the functional and non-functional requirements of the app. |
| Modelling | Using models that capture the structure, behavior, and functionality of the app. Then using provided tools to automatically generate code from the models. |
| Testing | Testing the generated code to ensure that it meets the requirements and functions correctly. |
| Deployment | Deploying the app to the target platform and audience, subsequent data management and analysis. |

Table 2

*Health Research Based Model Driven Development.*

| Stage | Description |
|---|---|
| Use case Models | To describe the various use cases or scenarios that the app will support. |
| Domain Models | Capture the key concepts and entities in the health research domain, such as patients, medical conditions, and treatments. |
| Process Models | Describe the workflows and processes that the app will support, such as data collection, analysis, and reporting. |

*Table 3*

*All Stakeholders involved in Physical Activity Research.*

| Role | Responsibilities |
| --- | --- |
| Researcher | Research Question; Data Analysis |
| Content Expert | Validate participant facing content |
| Software Developer | Design, develop, test, evaluate, maintain the Application |
| Participant | User Experience (UX) |

Figures

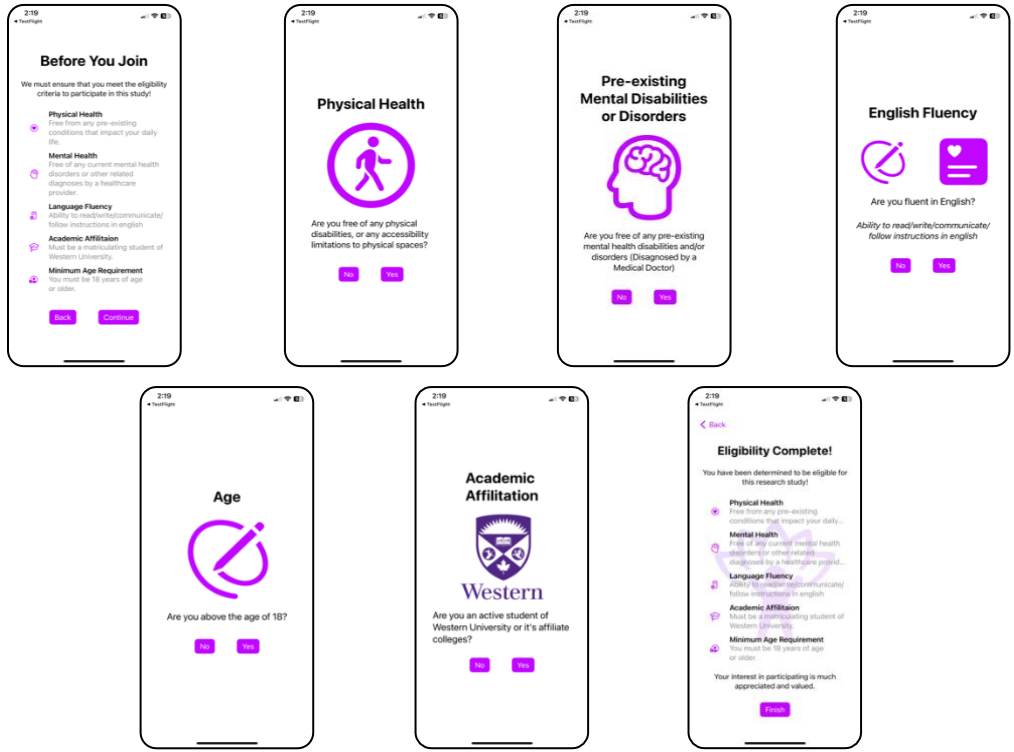

Figure *1*. Eligibility Process visualized in sequential order using visual and written means.

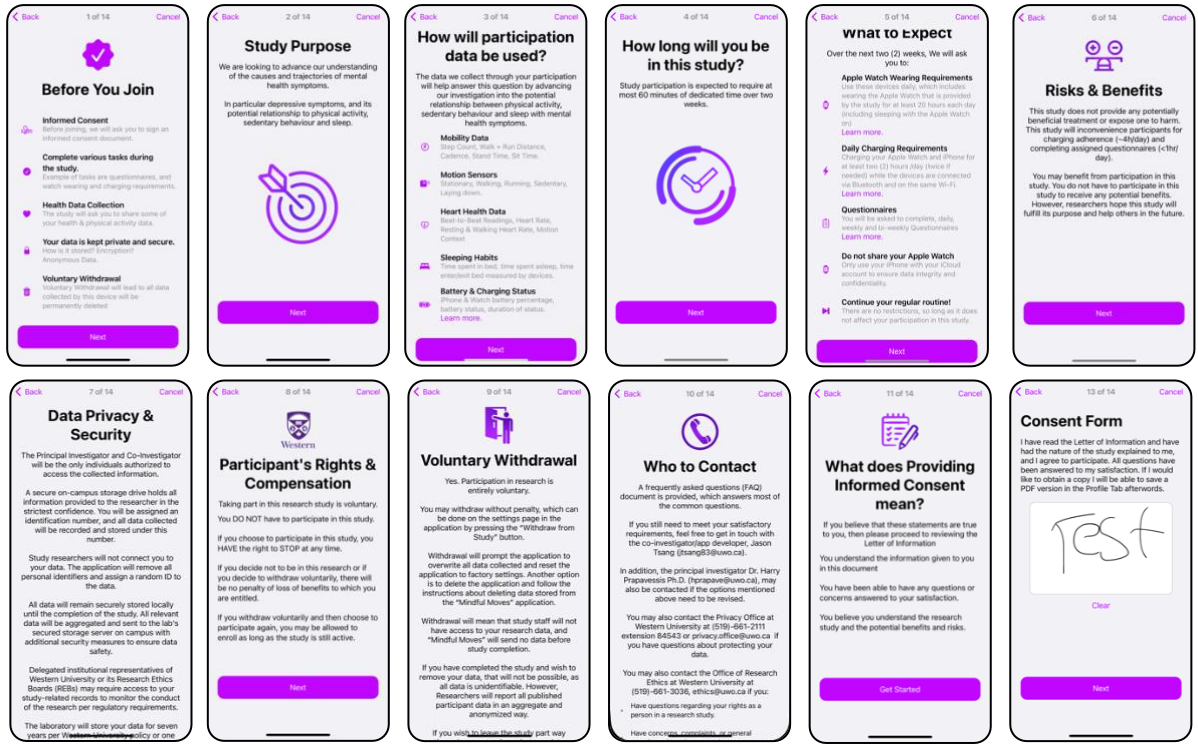

Figure *2*. Informed Consent Process visualized in sequential order using visual and written means.

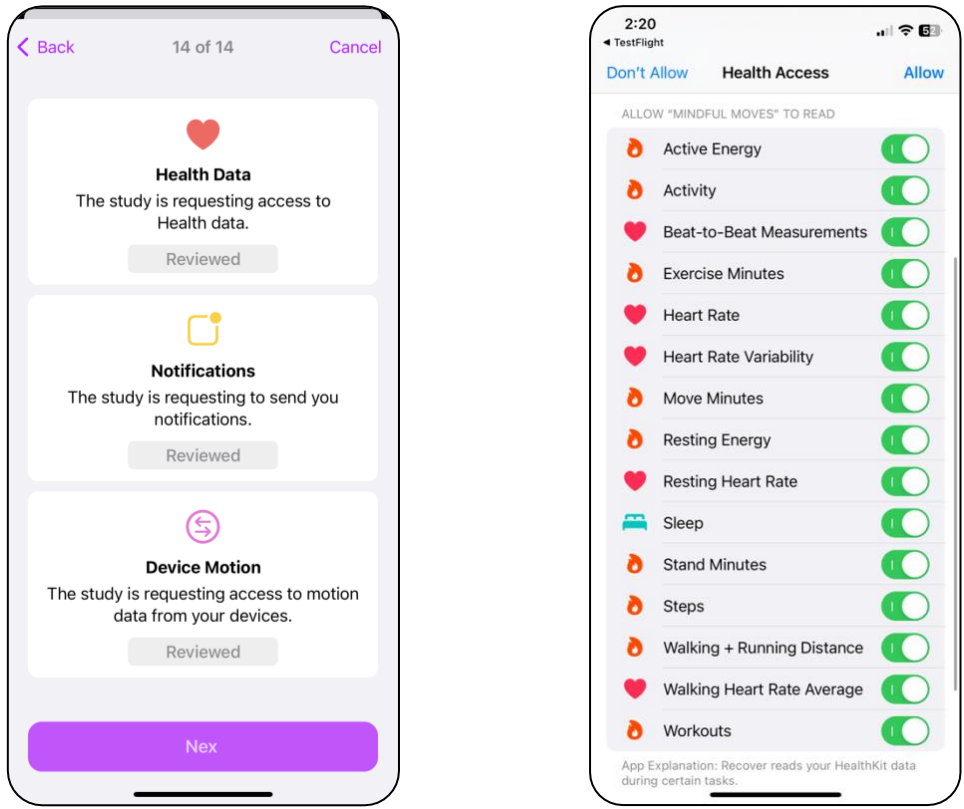

*Figure 3*. Permission gathering and disclosures are collected with permission needed for each individual component.

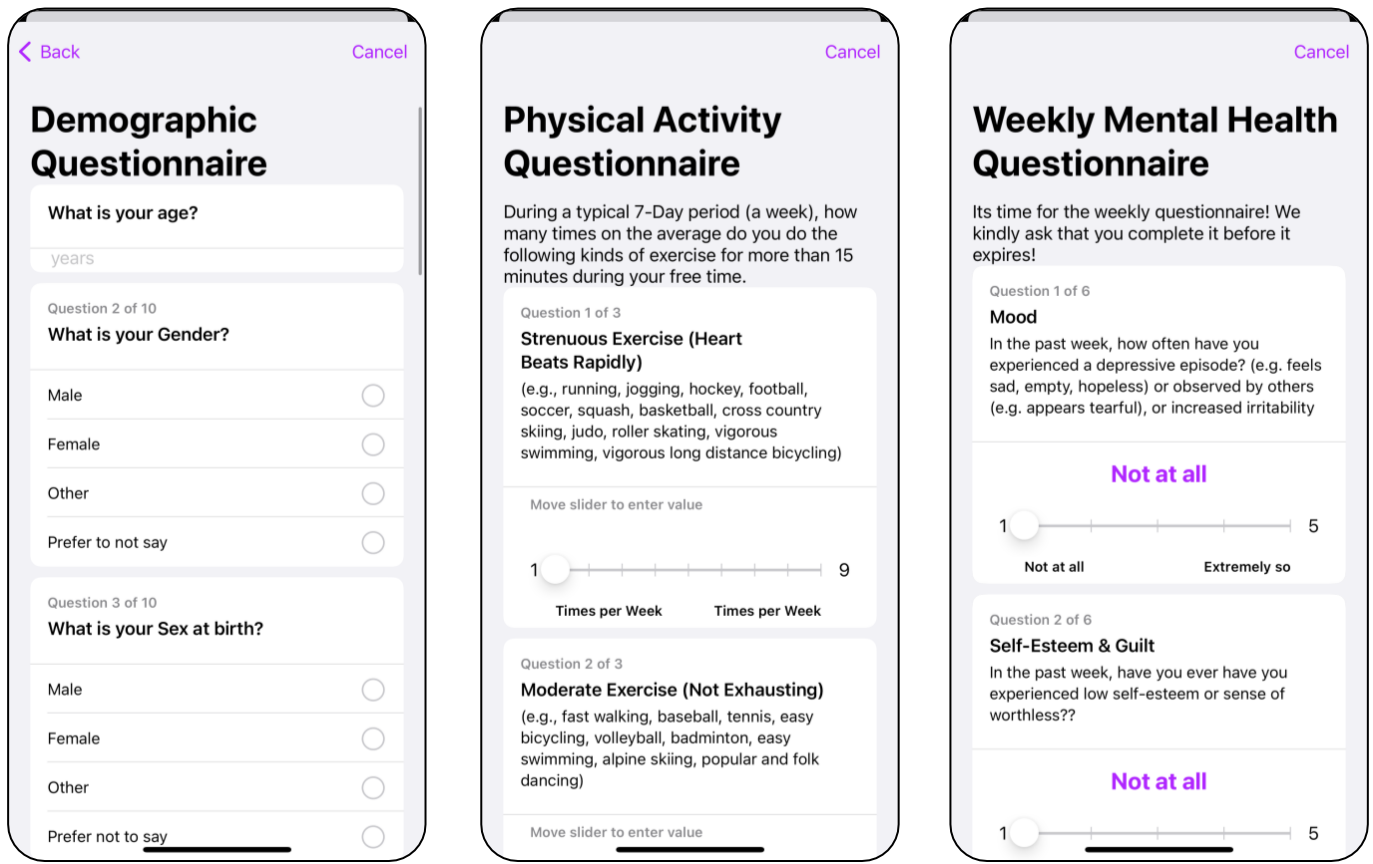

*Figure 4.* Examples of questionnaires created and administered directly on the application. All questionnaires were created using the survey's module.

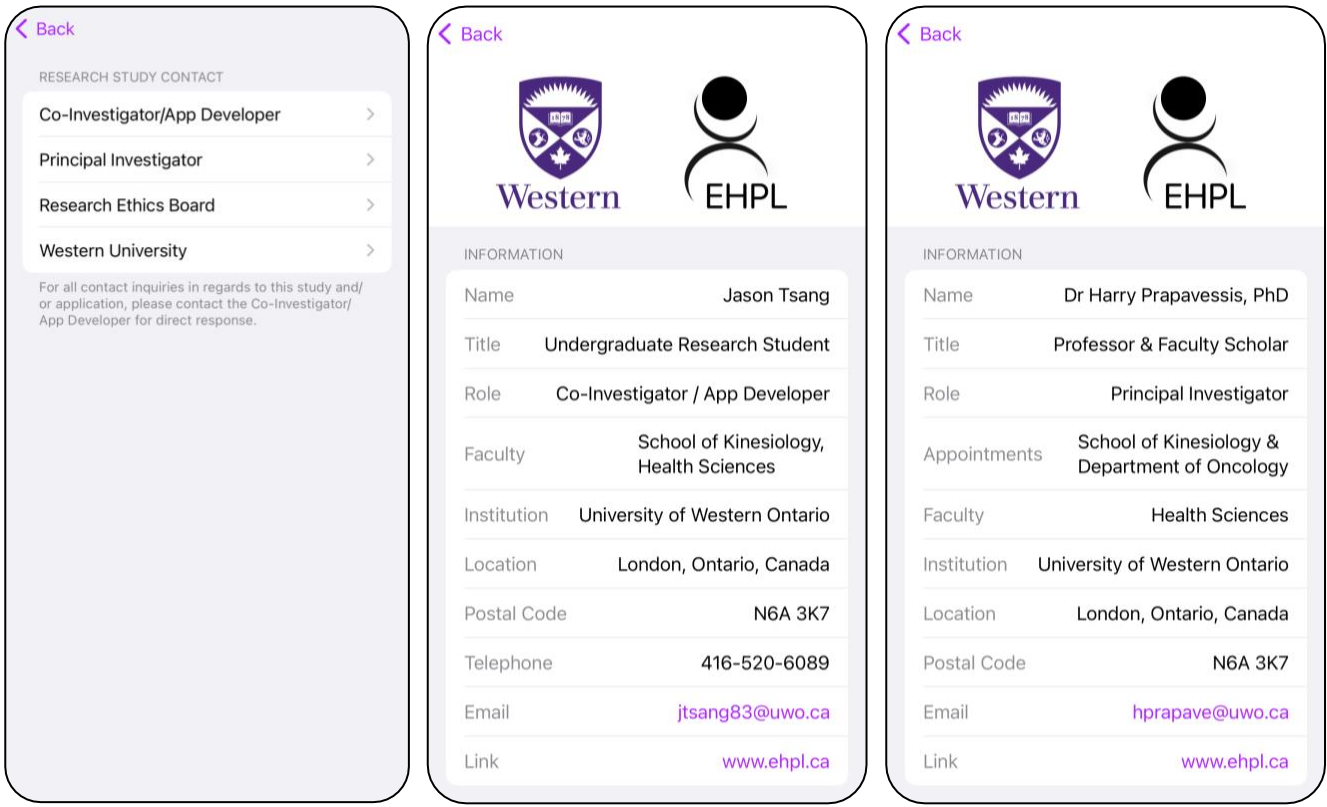

*Figure 5.* Visualization of point of contacts in the research-based application.

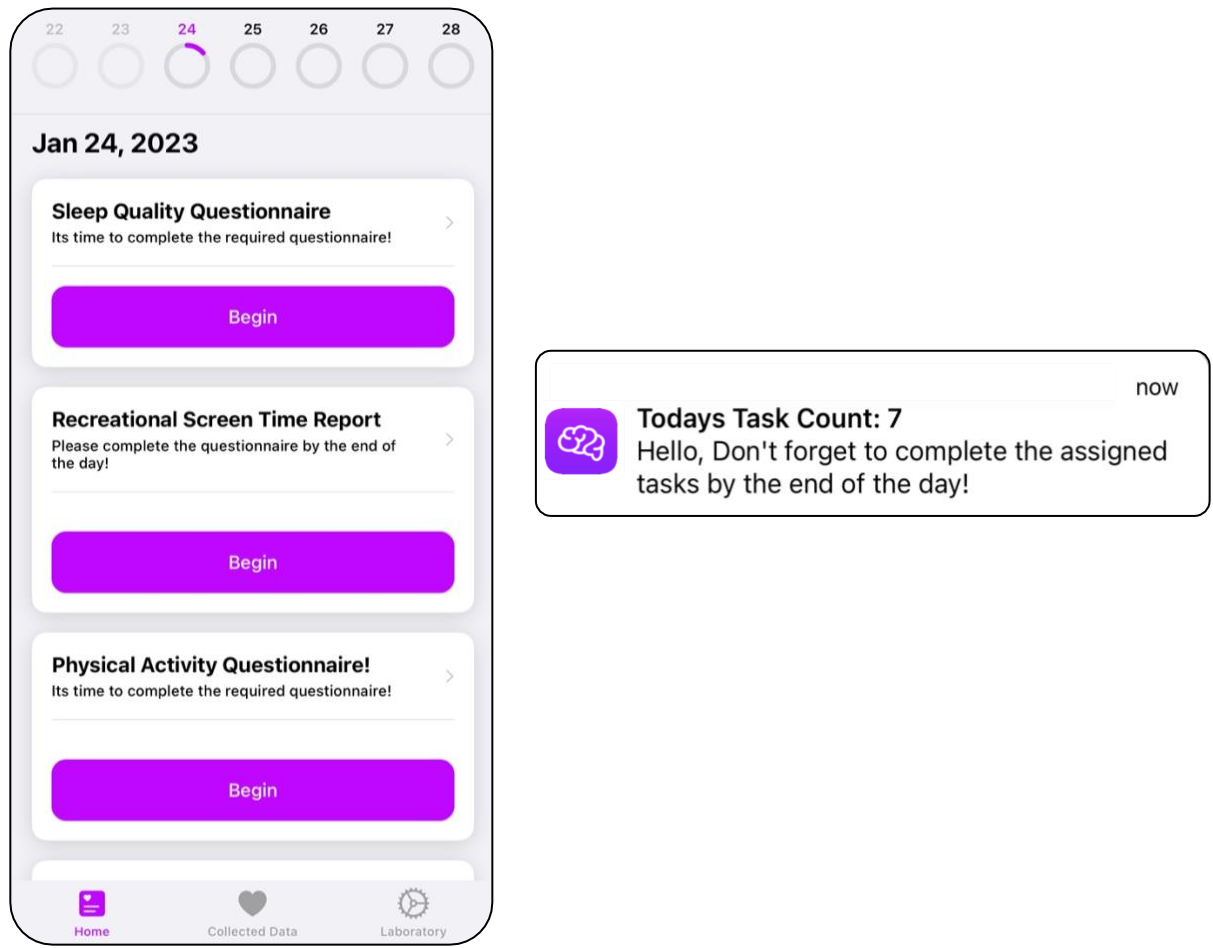

*Figure 6*. Features such as reminders, progress tracking, and a screen was created to motivate users and provide them with a way to track their progress in the study.

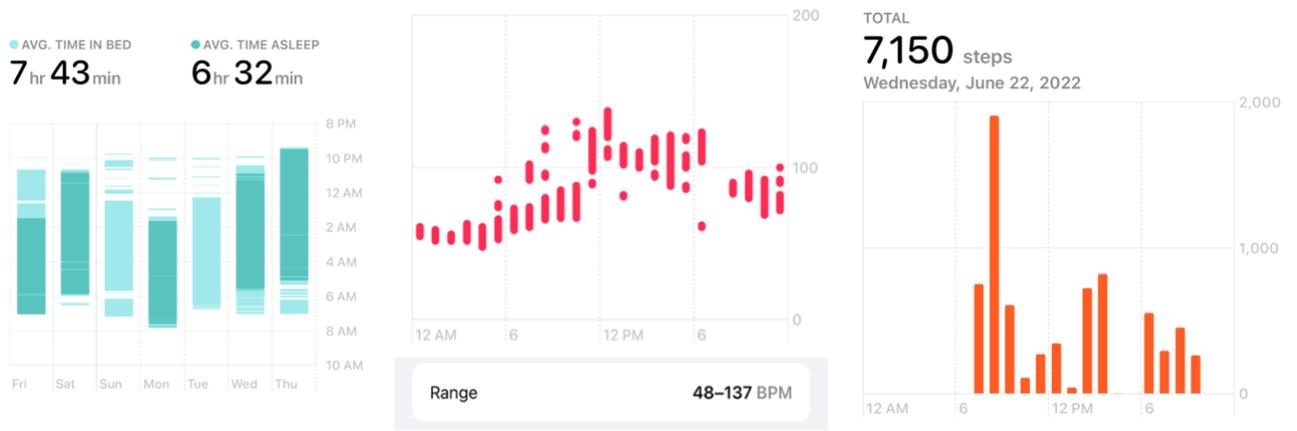

*Figure 7*. Visual representation of examples of collected biometric data to be analyzed by the researcher.